\documentclass[9pt,conference]{IEEEtran}

\usepackage{amsmath,amsfonts}
\usepackage{algorithmic}
\usepackage{algorithm}
\usepackage{array}
\usepackage{textcomp}
\usepackage{stfloats}
\usepackage{url}

\usepackage{verbatim}
\usepackage{graphicx}
\usepackage{cite}
\usepackage{xcolor}
\usepackage{booktabs}
\usepackage{tabularx}
\usepackage{multirow}
\usepackage{makecell}
\usepackage{subcaption}
\usepackage{float}

\begin{document}

\title{Cascade: An Application Pipelining Toolkit for Coarse-Grained Reconfigurable Arrays}

\author{\IEEEauthorblockN{\vspace*{-0.6cm}\\Jackson Melchert, Yuchen Mei, Kalhan Koul, Qiaoyi Liu, Mark Horowitz, Priyanka Raina}
\IEEEauthorblockA{Stanford University, Stanford, CA\vspace*{-0.3cm}}}

\date{}

\maketitle

\begin{abstract}

While coarse-grained reconfigurable arrays (CGRAs) have emerged as promising programmable accelerator architectures, pipelining applications running on CGRAs is required to ensure high maximum clock frequencies. Current CGRA compilers either lack pipelining techniques resulting in low performance or perform exhaustive pipelining resulting in high energy and resource consumption. We introduce Cascade, an application pipelining toolkit for CGRAs, including a CGRA application frequency model, automated pipelining techniques for CGRA application compilers that work with both dense and sparse applications, and hardware optimizations for improving application frequency. Cascade enables 7 - 34$\times$ lower critical path delays and 7 - 190$\times$ lower EDP across a variety of dense image processing and machine learning workloads, and 2 - 4.4$\times$ lower critical path delays and 1.5 - 4.2$\times$ lower EDP on sparse workloads, compared to a compiler without pipelining.

\end{abstract}
\section{Introduction}



Coarse-grained reconfigurable arrays (CGRAs) have been widely studied in recent years as efficient and performant configurable accelerator architectures. A CGRA can achieve better performance and energy efficiency than an FPGA, while maintaining much more flexibility than an application-specific integrated circuit (ASIC). To achieve performance and energy-delay product (EDP) that is competitive with ASICs, CGRAs need to execute applications at high clock frequencies. 

There are several classes of CGRA architectures that each have their own solution to the challenge of achieving high application clock frequencies. The first class is CGRAs that have  exhaustively pipelined interconnects \cite{epimap,dresc,modulo-scheduling,constraint-centric-scheduling}. These architectures register data at every interconnect hop, or boundary between tiles. Exhaustive pipelining introduces several challenges, mainly that the path a piece of data takes through the interconnect will determine the cycle at which the data arrives. In some compilers, the place and route tools are tightly coupled with the scheduling tools and paths are balanced to ensure that the application executes correctly \cite{dresc}. In other compilers, the targeted CGRA contains a network-on-chip (NoC) which handles dynamic network delays \cite{sara}. Both of these approaches are expensive; they either introduce additional hardware or result in high register usage.

The second class consists of CGRAs that are not exhaustively pipelined \cite{hycube} and have configurable interconnect registers similar to FPGAs \cite{sb_regs,hyperflex}. These architectures have the ability to route data along multiple hops on the interconnect in a single cycle. These CGRAs have the benefit of having a lower hardware cost than NoCs and more flexibility when choosing routes during place-and-route (PnR) \cite{hycube}. In this style of CGRA, the critical path while running an application is determined by the configuration of the array. A compiler that routes data with short net lengths will allow the CGRA to run at higher frequencies, improving performance and EDP. As the size of the CGRA increases, minimizing the critical path of applications becomes more difficult.

Minimizing critical paths is a widely studied problem across design tools and compilers for ASICs and FPGAs. However, CGRAs have considerations that makes a dedicated toolkit for pipelining CGRA applications necessary. First, CGRAs have word-level interconnects and processing elements (PEs). These elements have considerable delay, meaning that a combinational path in the application can have only a few elements along it before its delay severely limits the clock frequency. Additionally, there are fewer pipelining resources on CGRAs compared to ASICs and FPGAs. ASIC design tools have the freedom to insert registers wherever needed and FPGA arrays are much larger and contain many more registers than CGRAs. Finally, many CGRAs have memory units with in-built configurable address and schedule generation logic \cite{amber}; this scheduling logic can be programmed to accommodate new delays introduced during application pipelining, providing a unique opportunity to use new pipelining strategies that leverage this flexibility. 

In this paper, we introduce Cascade, an application pipelining toolkit for CGRAs. We target a class of CGRAs like \cite{amber}, which have large tile arrays, a configurable interconnect that allows for single-cycle multi-hop connections from any tile to any other tile, and configurable pipelining registers within every hop of the interconnect. To enable high maximum clock frequencies for applications running on such architectures, dedicated pipelining techniques are required.  

The major contributions of this paper are:

\begin{enumerate}
    \item A methodology for generating timing models of CGRAs.
    \item A static timing analysis tool that uses the timing model to determine the critical path of an application on a CGRA.
    \item Automated software pipelining techniques integrated into a CGRA application compiler, that apply to both dense and sparse workloads, along with a hardware optimization to further improve application frequency.
\end{enumerate}

Our evaluation shows that Cascade enables 7 - 34$\times$ lower critical path delays and 7 - 190$\times$ lower EDP across a variety of dense image processing and machine learning workloads, and 2 - 4.4$\times$ lower critical path delays and 1.5 - 4.2$\times$ lower EDP on sparse workloads, compared to a compiler without pipelining. 
\section{Related Work}

Application pipelining when targeting FPGAs and ASICs is a well studied problem; however, many of these techniques cannot be directly applied to CGRA applications or do not achieve the same level of improvement for CGRAs. Register retiming \cite{reg_retime,retiming} is a very common pipelining technique where pipelining register stages are added to a design and retimed into optimal positions. Register retiming can be applied to CGRA applications, however, limited register resources and long delays through PE tiles means that retiming is not as effective when targeting CGRA applications.


Post-placement retiming and pipelining has been applied in the past to both ASICs and FPGAs. \cite{post-pnr-regs} introduces post-placement retiming and register insertion for ASICs. The critical paths are identified post-placement and iteratively pipelined to enable higher maximum frequency. Similarly, Intel's Stratix 10 \cite{hyperflex} architecture includes pipelining registers in the interconnect and does post-place-and-route performance tuning including retiming and pipelining. This class of techniques can be applied to CGRA applications, although the limited register resources and long delays through PE tiles means that the algorithms used for ASICs and FPGAs do not directly apply to CGRA applications.

Current CGRA compilers either lack pipelining techniques resulting in low performance, or perform exhaustive pipelining resulting in high energy and resource consumption. Sara \cite{sara}, the compiler for the Plasticine architecture \cite{plasticine}, assumes an interconnect that is pipelined at every switch box. To accommodate this exhaustive pipelining, Plasticine has a NoC that handles dynamic network delays. This NoC and exhaustive pipelining results in high power consumption, \cite{hycube} reports that a NoC has a 28\% higher power cost than the alternative without exhaustive pipelining. Similarly, \cite{scalable-interconnects-compiler} introduces a compiler that targets a CGRA with a credit-based interconnect. 

Epimap \cite{epimap}, DRESC \cite{dresc}, and the compilers introduced in \cite{modulo-scheduling} and \cite{constraint-centric-scheduling} use exhaustive pipelining, where the paths in the mapped application need to be balanced to guarantee that the data arrives at each tile at the correct cycle. These compilers target CGRA architectures that do not have single-cycle multi-hop connections.

HyCUBE \cite{hycube} is a CGRA architecture that supports single-cycle multi-hop connections between tiles. This work acknowledges that this architecture may have longer critical paths than other types of interconnects, so they limit the number of hops a piece of data can make on the interconnect to one cycle. The HyCUBE array has only four rows and four columns, so the longest possible critical path is very short compared to our target CGRA architecture. 


\section{Background}

As background, we first summarize the class of CGRAs we target and describe algorithms for static timing analysis and branch delay matching that we build upon to create our pipelining flow.

\subsection{CGRA Architecture}

\begin{figure}
    \centering
    \includegraphics[width=0.37\textwidth]{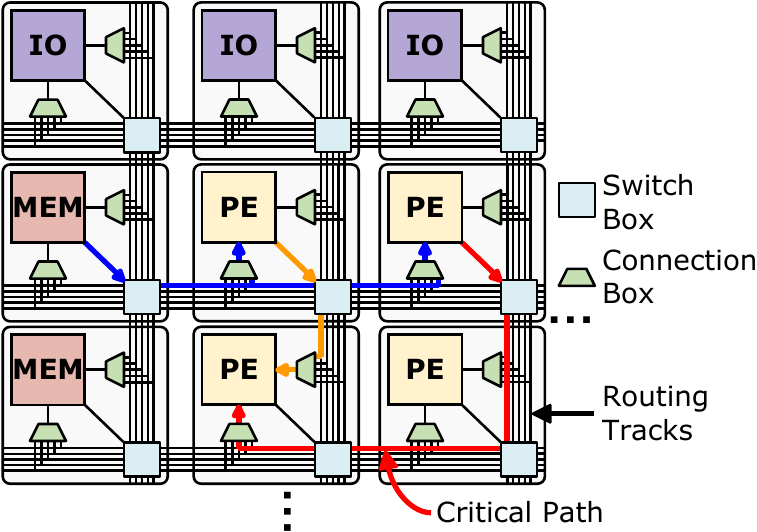} 
    \caption{CGRA architecture and application critical path model. This model uses a library of CGRA component delays and static timing analysis to calculate the critical path of a CGRA application and the maximum frequency that the application can be run at.}
    \label{fig:sta}
    \vspace{-0.4cm}
\end{figure}

CGRAs are typically composed of several processing element (PE) tiles, memory (MEM) tiles, and input/output (IO) tiles, arranged in a grid, as shown in Fig.~\ref{fig:sta}. The tiles communicate via a configurable interconnect, which is comprised of several horizontal and vertical routing tracks, connection boxes that bring inputs into the tiles from the routing tracks, and switch boxes that take outputs from the tiles and route them onto the routing tracks in different directions. In this paper, we primarily target CGRAs with large tile arrays (e.g. 512 tiles), a configurable interconnect that allows for single-cycle multi-hop connections from any tile to any other tile, and configurable pipelining registers within every switch box.

\begin{figure*}
    \centering
    \includegraphics[width=0.9\textwidth]{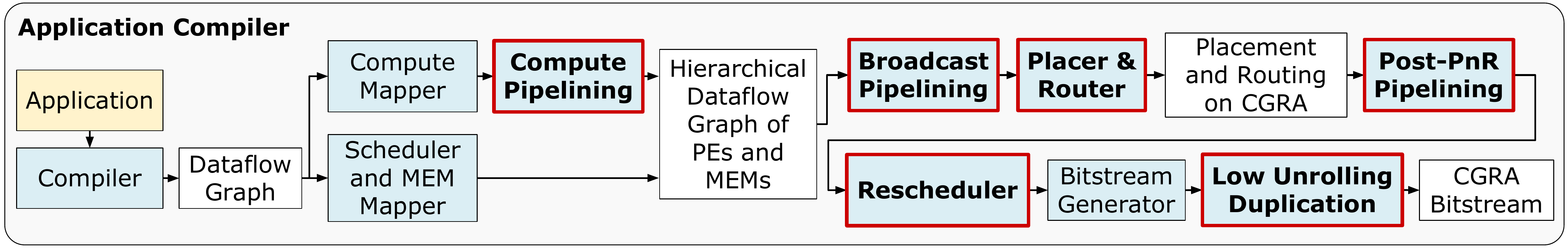} 
    \caption{CGRA application compiler that takes an application as an input and produces a CGRA bitstream. Intermediate representations are dataflow graphs. Cascade introduces all pipelining techniques shown in bold font with red borders that were not in the baseline compiler.}
    \label{fig:compiler}
    \vspace{-1em}
\end{figure*}

\subsection{Static Timing Analysis and Branch Delay Matching}
\label{sec:sta}

Static timing analysis (STA) is a standard technique for determining the critical path, that is the path with the maximum delay, through a circuit \cite{sta}. STA can be applied to any directed acyclic graph (DAG), and it involves iterating through every node in the DAG in reverse topological order and calculating the arrival time at each node. 






When pipelining registers are added to an application DAG, we need to ensure the correct execution of the application through branch delay matching, which matches the cycle arrival times of every piece of data arriving at every functional element in the application. We use an algorithm similar to STA for branch delay matching, but instead of using the delay through each hardware element, we instead use the number of cycles each node takes to generate an output. If we find a node that has more than one unique arrival time, we must insert registers to ensure correct application execution.


    
            


\subsection{Static Scheduling of CGRA Applications}
Many image processing and machine learning applications have statically analyzable access patterns. This characteristic enables CGRA compilers to statically schedule such applications (e.g., all memory accesses can be scheduled at compile time). The application scheduling process turns the multidimensional loop statements in these applications into operations (load/store) executed on the memory tiles. 
The CGRA compiler we build upon \cite{tecs} generates a cycle-accurate schedule which gives each statement in the application's iteration domain a one-dimensional timestamp, which represents the hardware's runtime behavior and enables pipelined parallelism. This static schedule overlaps data transfer with computation and extracts data reuse to reduce latency and improve energy efficiency. 




\section{Modeling CGRA Application Frequency}
\label{sec:compiler}
We first describe our CGRA application compiler toolchain, the attributes of the CGRA hardware that contribute to the path delays, and our static timing analysis tool specifically for CGRA applications. We develop our pipelining framework on top of an existing open-source agile hardware design toolchain that encompasses application specification, scheduling, mapping, place and route, and bitstream generation~\cite{tecs}, as illustrated in Fig.~\ref{fig:compiler}. 


\begin{figure}
    \centering
    \includegraphics[width=0.4\textwidth]{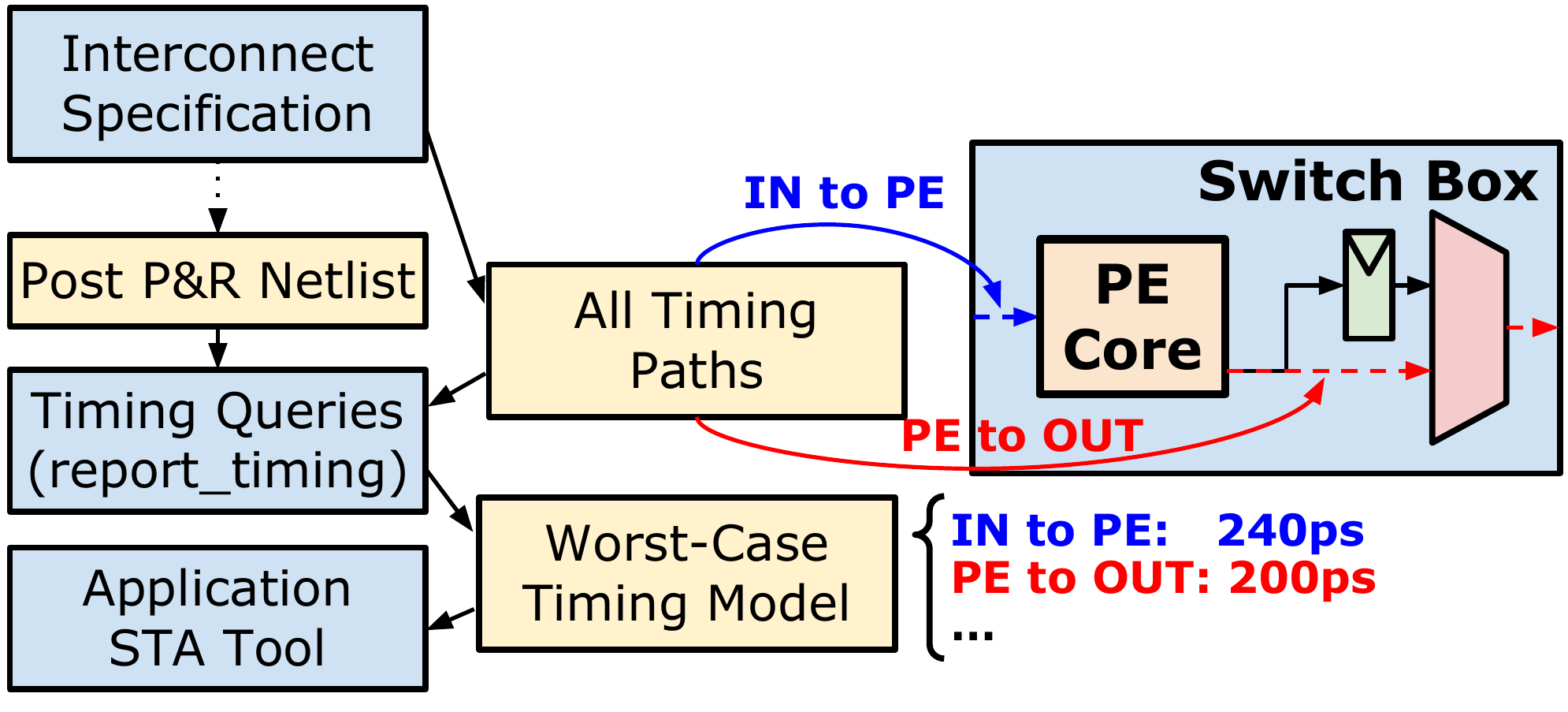} 
    \caption{Given an interconnect specification written in Canal \cite{tecs}, we automatically generate all paths of interest in a PE or memory tile. Using a commercial ASIC static timing analysis tool we can find the worst-case delay of these paths and use them in the application STA.}
    \label{fig:methodology}
    \vspace{-0.3cm}
\end{figure}

\subsection{Methodology for Generating an Application Timing Model}




Generally, a path within the CGRA (rather than a path at the boundary between the CGRA and the memory system) will determine the maximum frequency of the accelerator. There are two components which have large contributions to path delay. First, there is significant delay associated with the functionality within a tile, which we call a core. For example, the PE tile core will include an ALU, and the memory tile core will include SRAMs. Second, on the CGRA, we can configure the interconnect to route any piece of data from one location on the array to another. By default, this path will have no registers along it, and its length will be determined by how many tiles it passes through, or ``hops" on the interconnect. 

On real hardware, all hops on the interconnect will not have the same delay. A memory tile has a much larger footprint than a PE tile, so traveling from one side of a memory tile to the other takes longer than the same path through a PE tile. We also find that the lengths of the wires going in one direction through a tile are not the same as those going in another direction. Finally, we also have to consider the possibility that the clock does not arrive at the same time at each tile due to clock skew. 


We propose a methodology for generating a timing model of a CGRA in an automated fashion, shown in Fig.~\ref{fig:methodology}. This model contains information about all significant delays in the CGRA, and is used later for application STA. To generate the model, we use an interconnect specification expressed in Canal \cite{tecs}. Canal gives us the ability to describe a wide range of CGRA interconnect topologies with a graph representation. Additionally, Canal automatically generates the interconnect RTL, place and route collateral, and bitstream configuration. We then add the ability to enumerate all possible data and clock paths at the tile level that have significant delays using the internal interconnect graph representation. Canal automatically generates the start and end points of these paths in the corresponding RTL representation. We then estimate the delays of these paths using a commercial static timing analysis tool running on the tile's post-place-and-route netlist with parasitics to accurately match the worst-case delays on real hardware. We use these worst-case delays in our application STA tool. 

\subsection{Static Timing Analysis of CGRA Applications}
\label{sec:app-sta}

We feed all the component delays described in the previous section into a static timing analysis tool for CGRA applications. The input to this tool is the application data-flow graph representation after the place-and-route stage of the compiler. As described in Section~\ref{sec:sta}, we use the static timing analysis algorithm to calculate the arrival time of each piece of data at every node in the data-flow graph. As shown in Fig.~\ref{fig:sta}, the maximum of these arrival times determines the resulting critical path and the maximum frequency of the application. We can use the STA tool to make pipelining decisions, for example if we have an application that we attempt to pipeline, the STA tool will let us know if we decreased the critical path or increased it. We evaluate the accuracy of this STA tool in the results section.

\section{Software Pipelining Techniques}
\label{sec:software}

In this section we describe automated pipelining techniques implemented in the application compiler to improve the critical path of CGRA applications. These techniques are summarized in Fig.~\ref{fig:compiler}. Each of these techniques solves a particular aspect of CGRA application pipelining, and when combined they achieve short critical paths and fast runtime.

\subsection{Compute Pipelining}

\begin{figure}
\centering
\begin{subfigure}[t]{0.4\linewidth}
    \centering
    \includegraphics[width=\linewidth]{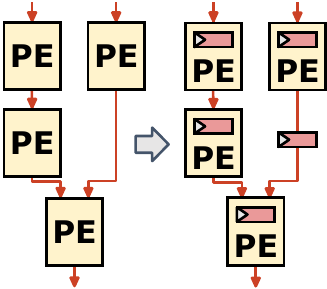}
\end{subfigure}\hfil
\begin{subfigure}[t]{0.5\linewidth}
    \centering
    \includegraphics[width=\linewidth]{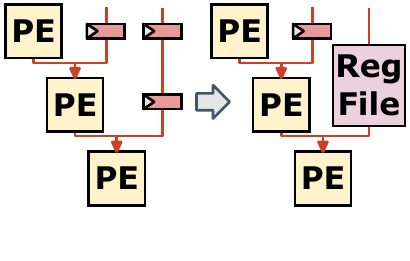}
 \end{subfigure}
 \caption{Left: Compute pipelining where registers at the inputs of PEs are enabled and branch delay matching is performed. Right: Register chains are transformed into register files configured as variable length shift registers.}
 \label{fig:compute-pipelining}
 \vspace{-0.5cm}
\end{figure}



During the compute mapping stage of the application compiler, we translate a DAG of primitive operations into a DAG of PEs. In our target CGRA architecture, each PE has configurable registers at the inputs of the PE. These registers can be either enabled or bypassed. 
The compute pipelining stage of the automated pipelining tool will turn on every available PE input register and then perform branch delay matching to ensure that the application compute kernels maintain their functionality, as shown in Fig.~\ref{fig:compute-pipelining}. 

Compute mapping may introduce a significant number of pipelining registers into an application. The CGRA has limited resources, including a limited number of available registers on the configurable interconnect. In some cases, many pipelining registers might exist in a chain after branch delay matching. Many CGRA architectures, including our target architecture, include register files throughout the tile array. In these cases we utilize register files in PE tiles to act as variable length shift registers to eliminate these long chains of registers, thereby freeing up resources. 
In Fig.~\ref{fig:compute-pipelining}, we show an example of this transformation for a chain of two registers. This transformation is applied to every $N$ chain of registers, where $N$ is a hyperparameter.

Compute pipelining is a technique applied on the abstract compute kernels of an application. To further improve application critical path lengths, we must use techniques focused on minimizing the actual wire delays present in mapped applications running on the CGRA.

\subsection{Broadcast Signal Pipelining}
\label{sec:broadcast}

After compute pipelining, we ran the application STA tool on a suite of benchmark applications and noticed that path delays on the configurable interconnect dominate the critical path in all applications. In our target CGRA architecture, the delay through a PE tile is a maximum of 0.7ns, while the delay through one switch box is about 0.14ns. If every PE is pipelined during compute pipelining, then the path delays in an application will become dominant after five hops on the interconnect.


We notice that certain paths that have one source and many destinations end up being routed inefficiently on the CGRA, and typically have many more than five hops on the interconnect. In our benchmark suite, every application has a broadcast path, and after applying compute pipelining, these become the limiting factor.


To solve this problem, we can specifically pipeline broadcast paths and use a tree structure to ensure that the maximum wirelength is minimized. There is a trade-off between number of registers added and critical path length, so the parameters of this transformation pass (number of tree levels, maximum number of pipeline registers, etc.) can be adjusted.

\subsection{Placement Algorithm Cost Function Optimization}

The placement and routing of an application on the CGRA is important when trying to minimize the critical path. After enabling compute pipelining and broadcast signal pipelining, we noticed that long routes that start at one side of the array and end at the other side were limiting the maximum frequency of the application. These long routes could be mitigated if the placement algorithm placed the source and destination closer together.

Our application compiler place and routes the application onto the CGRA array using a simulated annealing based placement algorithm and an iteration-based routing algorithm \cite{tecs}. 
The cost function for detailed placement (see Equation \ref{eqn:detailedplacement-function}) is calculated by summing the half-perimeter wire length (HPWL) cost for each net and adding an additional term to penalize pass-through tiles. Pass-through tiles are those that are only used for routing. $\gamma$ is a hyperparamter that adjusts how much penalty pass-through tiles have.

To improve the critical paths of applications that are placed and routed using this technique, we add another hyperparameter $\alpha$ that penalizes longer routes, similar to the criticality exponent introduced in \cite{hwpl}. Typical placement algorithms do not use this parameter, and therefore minimize total wirelength. By introducing $\alpha$, we can adjust how much long routes are penalized. A higher value will mean that long routes cost much more than shorter routes. 

\vspace{-0.4cm}
\begin{equation}
\label{eqn:detailedplacement-function}
\text{Cost}_{net} = (\text{HPWL}_{net} + (\gamma \times \text{Area}_{pass-through}))^\alpha
\end{equation}


\subsection{Post-Place-and-Route Pipelining}

\begin{figure}
    \centering
    \includegraphics[width=0.35\textwidth]{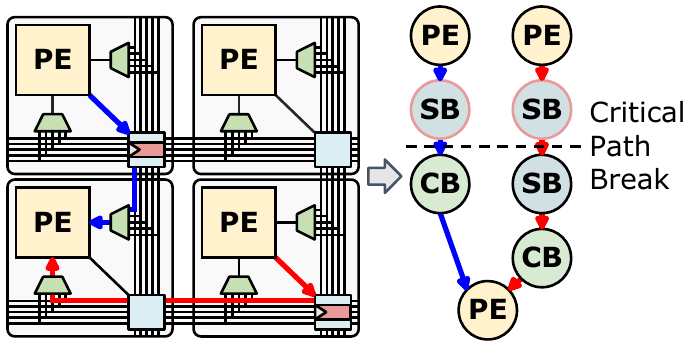} 
    \caption{Post-PnR pipelining takes the place and route result represented as a data-flow graph and performs STA to identify the critical path. This path is then broken by enabling registers in the switch box (SB), and the graph is branch delay matched.}
    \label{fig:post-pnr-pipe}
    \vspace{-0.6cm}
\end{figure}

When examining the final critical paths in our application after applying each technique described above, we determined that the most effective way to break those paths would be to directly insert registers into the place-and-routed result, similar to the technique for physical design register placement in \cite{post-pnr-regs}, and FPGAs in \cite{hyperflex}.

After place and route (PnR) is complete, we know exactly where each tile will be placed in the array and where the nets will be routed. 
During post-PnR pipelining, we iteratively identify the critical path and insert pipelining registers to break it.

The interconnect of our CGRA architecture has configurable pipelining registers within every switchbox of the array. These pipelining registers exist on every 16-bit and 1-bit track going out of the switchbox in each of the four directions. After place and route has completed, we can modify the configuration of the CGRA to turn on individual pipelining registers. This affects the execution of the application, so we must do branch delay matching in order to maintain application functionality.

For example, in Fig.~\ref{fig:post-pnr-pipe}, the red path is the longest in the application, so it is the critical path. During post-PnR pipelining, we break this path by turning on the pipelining register in the bottom-right switch box. Next, we run branch delay matching to ensure that the application still functions correctly, inserting pipelining registers to balance any paths if needed. In this example, we need to break the blue path to balance the arrival times of the two pieces of data at the final PE. Then, we analyze the application again and repeat the process until we cannot break any more paths.

\subsection{Low Unrolling Duplication}

Next, we describe low unrolling duplication, a transformation on the application specification and the final configuration bitstream. Image processing and machine learning applications are typically unrolled on hardware accelerators. This allows for parallelism, producing more than one output pixel per cycle, and is critical to achieving fast application runtimes. 
We found that running applications with no unrolling and performing place and route on smaller CGRAs often lead to much shorter critical paths. The configuration of the tiles and interconnect is then duplicated across the array ``unrolling" the application in the exact same way every time. This optimization allows the place and route tool to solve a much smaller problem while maintaining all of the benefits of unrolling.

\subsection{Modifying Application Schedules Post Pipelining}
When adding pipeline registers to a statically scheduled CGRA application, the schedule needs to be updated to reflect any changes to the compute latencies. 
Because the topology of a mapped application graph will not change with different compute latencies, in the first round of application compilation we set all computation latencies to 0. The mapped application graph will be sent to the downstream toolchain to give us a precise compute latency through the concrete PE DAG post-PnR.
In the next round of scheduling, post-pipelining, the compute latencies will be updated based on the PnR result, and we rerun the flow to get the final pipelined and scheduled application.


\section{Hardware Pipelining Techniques}
\label{sec:hardware}




With broadcast signal pipelining, expensive broadcast routes causing long critical paths can be shortened. However, for broadcast paths that have many destinations, the number of registers that need to be placed on the CGRA is very large.
Static scheduling orchestrates exactly where each piece of data is at every cycle during program execution. In every PE tile and memory tile there are memories, and controllers that read and write to the memories based on that static schedule. These elements need to be synchronized at the beginning of application execution, and that is done using a global flush signal.

This flush signal is a broadcast path with one source and potentially hundreds of destinations, depending on the size of the application. Pipelining this broadcast path using the technique described in Section~\ref{sec:broadcast} is not feasible for these applications, so we hardened this signal. Instead of routing this signal in the configurable interconnect, we directly connect it to every tile outside of the interconnect. This signal is routed from the top of the array to the bottom, running down each column. 



\section{Pipelining Sparse Applications}
\label{sec:sparse}

So far we have focused our discussion on statically scheduled applications. However, not all applications can be completely statically scheduled. In particular, sparse applications may have data-dependent memory accesses. When targeting a sparse CGRA application, we need to include several additional considerations.

Sparse applications typically use a ready-valid interface between all stages of an application. If a piece of data is routed from $Tile\ A$ to $Tile\ B$, a valid signal will be routed in the exact same way, going through the same switch boxes and connection boxes. A ready signal will be routed in the same way but in the opposite direction. Each of these wires travel the same distance as the original data signal.

If we identify a route in a sparse application as the critical path, we cannot increase application frequency by adding a register to the data signal alone, we need to register the ready and the valid signals as well. However, na\"{\i}vely adding pipelining registers to all three signals would break the ready-valid interface, which relies on single-cycle communication of when a component is ready to receive another piece of data. 
Therefore, all the techniques from Section~\ref{sec:software}, except placement cost function optimization and low unrolling duplication, require a small change. The change entails inserting FIFOs instead of registers to break the long data, ready, and valid paths together. These FIFOs include logic for handling the ready and valid signals correctly, and allow us apply our pipelining flow to sparse applications.
\section{Results}

We first evaluate the prediction accuracy of the application STA model using SDF-annotated gate-level simulations. Then, we assess the impact of each of the software and hardware pipelining techniques on maximum frequency, runtime, and power of several dense and sparse applications. We evaluate these techniques within the application compiler shown in Fig.~\ref{fig:compiler}. The CGRA architecture that we use is a $32\times16$ array with 384 PE tiles and 128 MEM tiles. We perform physical design of the CGRA in GlobalFoundries 12 nm technology. Fig.~\ref{fig:sta-vs-sdf} shows the final layout of the CGRA, along with a global buffer for storing input, output and intermediate data from the applications.

\subsection{Evaluating the Application STA Model}

We evaluate the application STA model to ensure that the critical path derived from the model matches real hardware. We use a standard delay format (SDF)-annotated gate-level simulation of the CGRA to search for the fastest clock period of applications with different pipelining techniques. The simulation is performed on a post-place-and-route version of the netlist and includes both gate delays and wire delays. The search granularity is 0.1ns.

\begin{figure}[b!]
    \centering
    \vspace{-0.4cm}
    \includegraphics[width=0.43\textwidth]{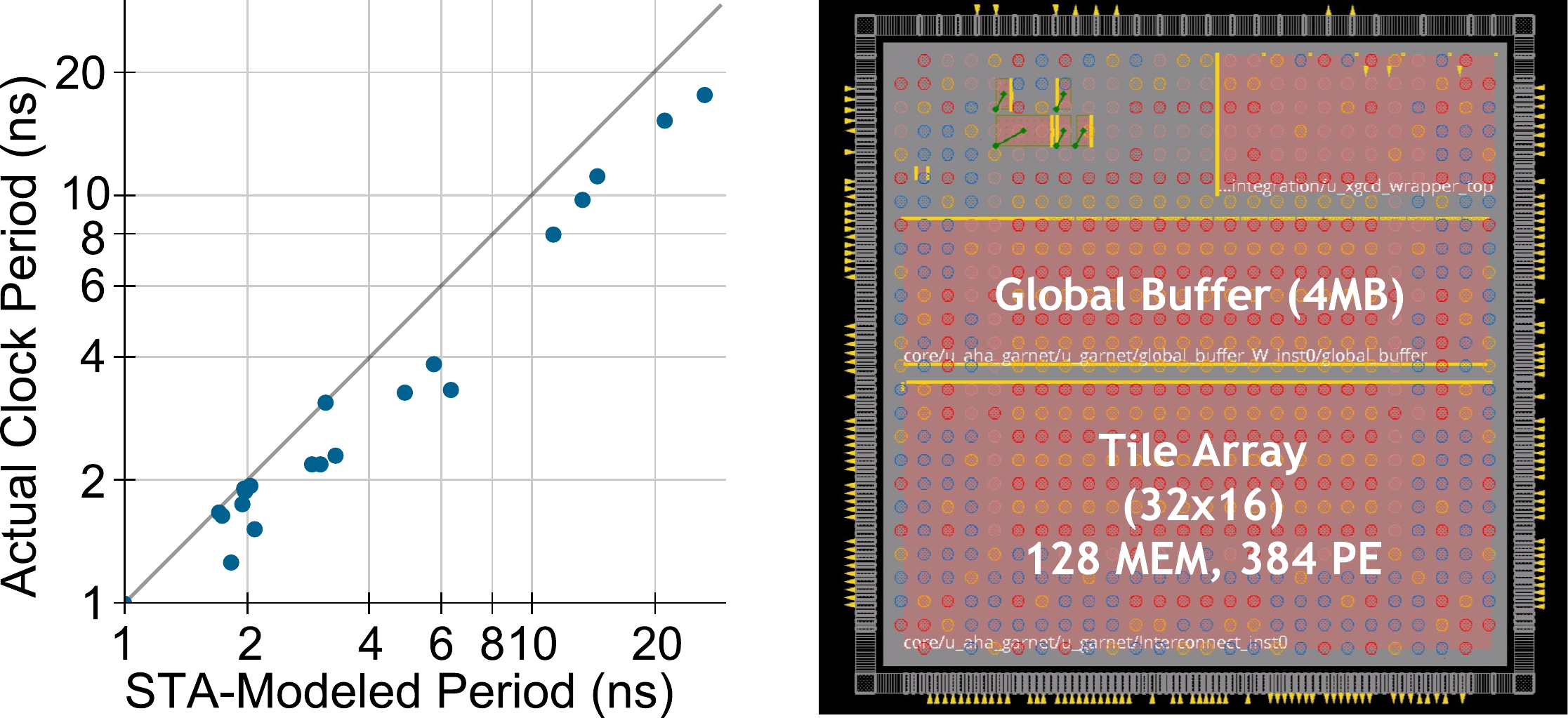} 
    \caption{Left: Evaluation of the STA critical path model. Each dot represents an application running at the frequency indicated by the vertical axis. The dot's horizontal value is the STA-modeled clock period. The gray line represents a perfect match. Right: Chip layout.}
    \label{fig:sta-vs-sdf}
    \vspace{-0.4cm}
\end{figure}

As shown in Fig.~\ref{fig:sta-vs-sdf}, the STA-modeled maximum clock period is generally higher than SDF-annotated gate-level simulation. This behavior is expected as we collected the worst-case path lengths when constructing this STA model. At clock frequencies above 500 MHz, which is the range we care the most about, the average error is 13\%. That is, the STA model has good accuracy for predicting the length of the critical path for applications running at high frequencies. Additionally, it must be noted that our model is pessimistic and provides a lower bound for the actual maximum frequency.

\subsection{Software Pipelining Techniques for Dense Applications}

\begin{figure} [!t]
    \vspace{-0.3cm}
    \centering
    \includegraphics[width=0.43\textwidth]{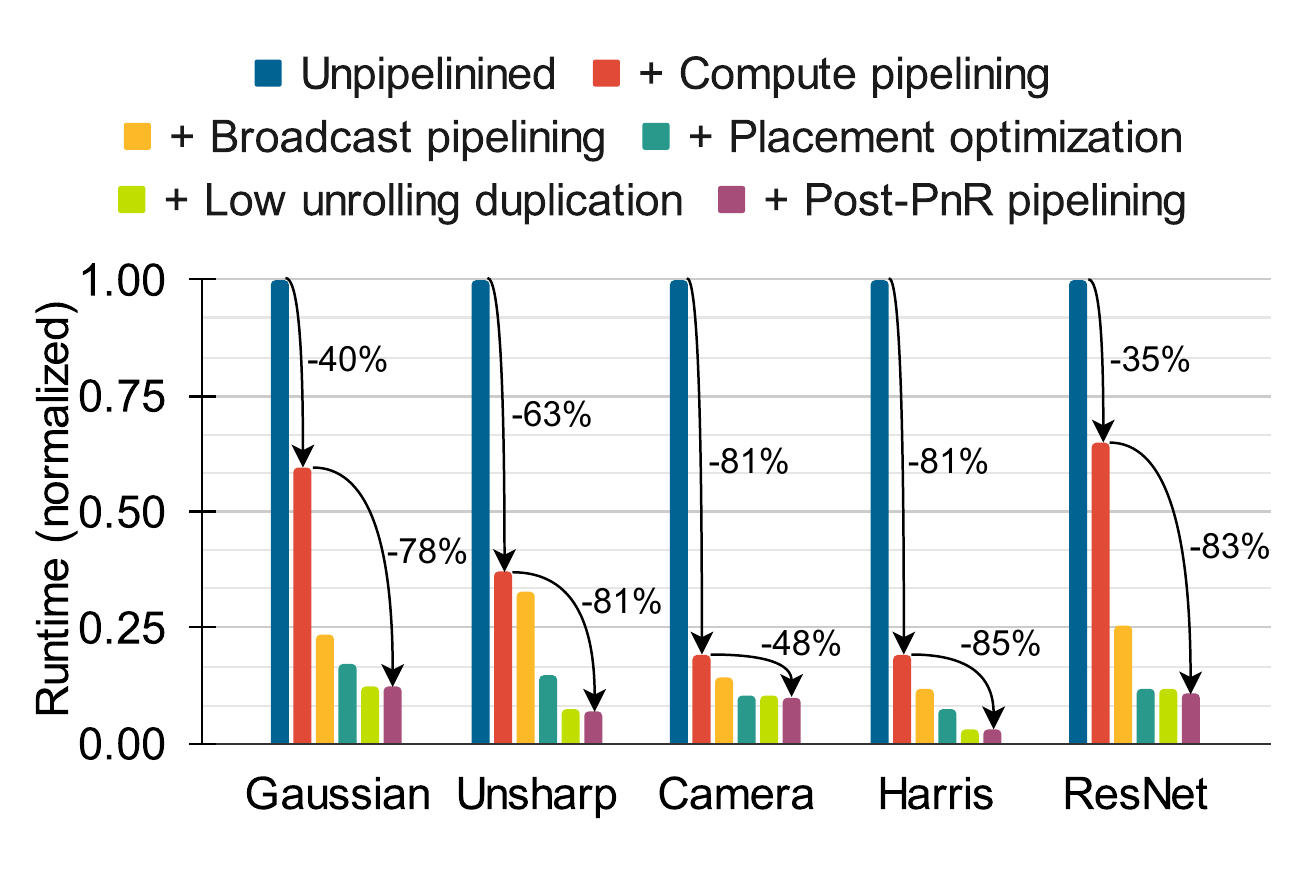} 
    \vspace{-0.5cm}
    \caption{Incremental effect of each software pipelining technique on dense applications.}
    \label{fig:software-pipelining}
    \vspace{-0.3cm}
\end{figure}

\begin{table}\footnotesize
    \caption{Frequency, runtime, and power comparison between unpipelined and pipelined versions of five dense applications. }\label{tab:software-pipelining}
    \vspace{-0.2cm}
    \begin{center}
        \renewcommand{\arraystretch}{1.2}
        \begin{tabular}{llccc}
        \hline\hline
        \multicolumn{2}{c}{Dense Application}          &   \makecell{Frequency \\ (MHz)} & \makecell{Runtime \\ (ms/frame)} & \makecell{Power \\ (mW)} \\
        \hline\hline
        \multirow{5}{*}{Unpipelined}&   Gaussian  &  103        &   22.6    &   156    \\
                                    &   Unsharp   &  66       &   21.4    &   139    \\
                                    &   Camera    &  47       &   28.3    &   318    \\
                                    &   Harris    &  30       &   70.6    &   85   \\
                                    &   ResNet    &  57       &   31.7    &   119    \\
        \hline
        \multirow{5}{*}{Pipelined}  &   Gaussian  &   610       &   3.66    &   841    \\
                                    &   Unsharp   &   532       &   1.99    &   903    \\
                                    &   Camera    &   457       &   2.96    &   678    \\
                                    &   Harris    &   571       &   1.90    &   614    \\
                                    &   ResNet    &   457       &   3.96    &   304    \\
        
        \hline\hline

        \end{tabular}
    \end{center}
    \vspace{-0.6cm}
\end{table}

\begin{figure} []
    \centering
    \vspace{-0.6cm}
    \includegraphics[width=0.44\textwidth]{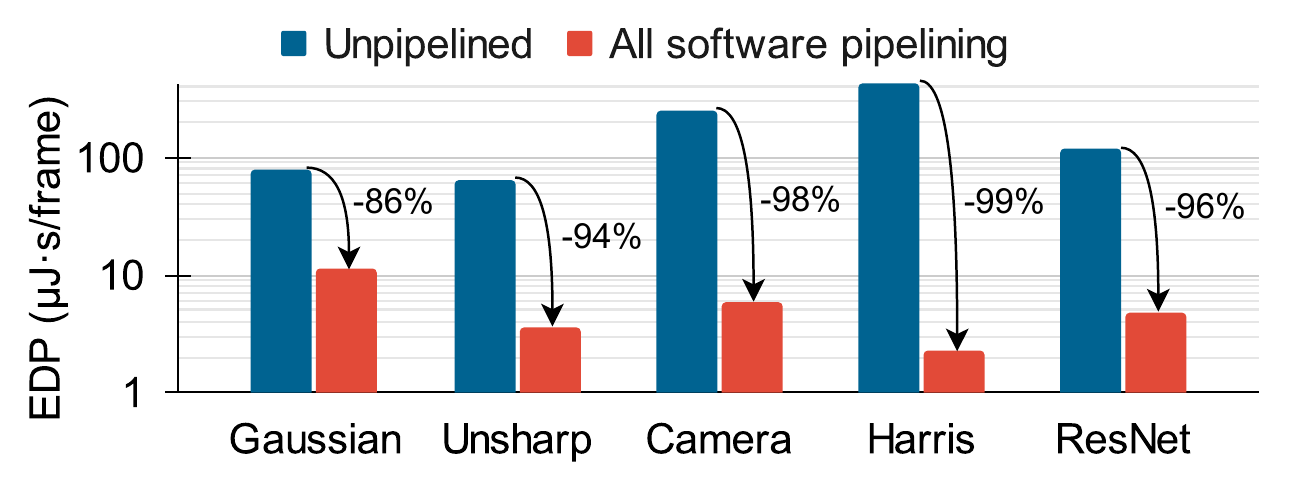} 
    \vspace{-0.2cm}
    \caption{The EDP of unpipelined applications versus applications with all software pipelining. The EDP decreases by 95\% on average.}
    \label{fig:EDP-dense}
    \vspace{-0.4cm}
\end{figure}

We analyze the incremental impact of each software pipelining technique from Section~\ref{sec:software} on the runtime of five dense applications from image processing and machine learning domains, which are also benchmarks in a previous CGRA work~\cite{tecs}. This is shown in Fig.~\ref{fig:software-pipelining} and Table~\ref{tab:software-pipelining}. The results in Fig.~\ref{fig:software-pipelining} are derived from our STA model, while the results in Table~\ref{tab:software-pipelining} are verified with SDF-annotated gate-level simulation. For image processing applications, the frame size is 6400$\times$4800 for Gaussian, 1536$\times$2560 for Unsharp, 2560$\times$1920 for Camera, and 1530$\times$2554 for Harris. For machine learning, ResNet refers to a single conv5\_x layer of ResNet-18. In these experiments, we have applied the hardware technique described in Section~\ref{sec:hardware}. Fig.~\ref{fig:EDP-dense} shows the impact of our pipelining flow on EDP. As shown in Table~\ref{tab:software-pipelining}, the software pipelining techniques achieve a 84 - 97\% decrease in runtime and a 86 - 99\% decrease in energy-delay product versus unpipelined implementations. Compute pipelining alone results in a 35 - 81\% reduction in runtime while the techniques applied during place and route result in an additional 48 - 85\% reduction in runtime.

\subsection{Hardware Pipelining Techniques for Dense Applications}

\begin{figure}[]
    \vspace{-0.3cm}
    \centering
    \includegraphics[width=0.44\textwidth]{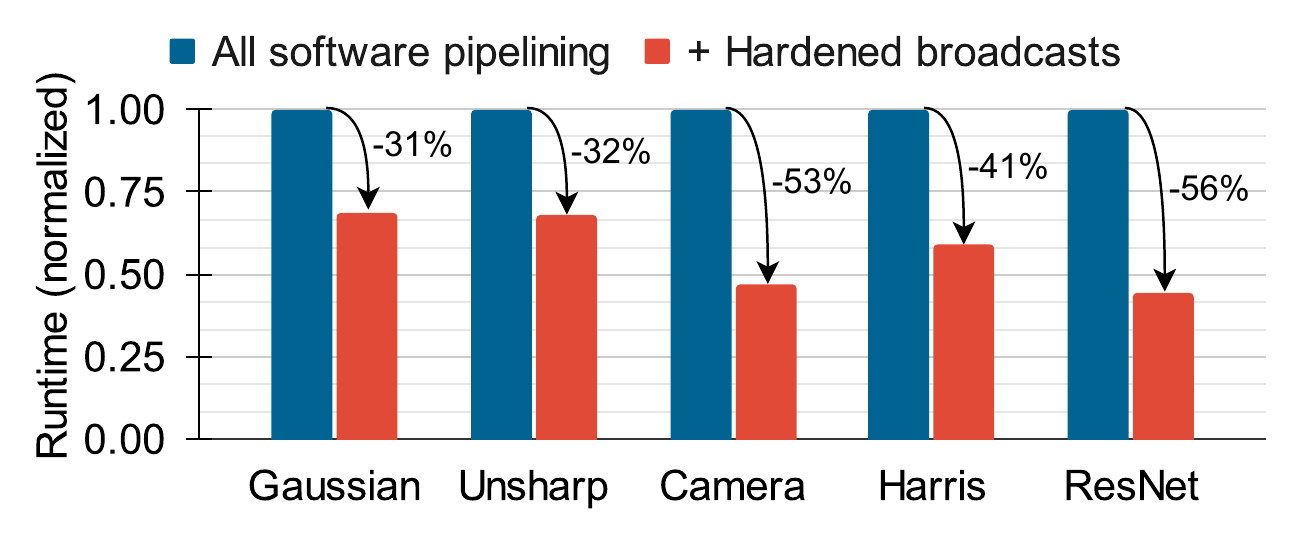} 
    \vspace{-0.2cm}
    \caption{Hardening of some broadcast signals that reach many tiles (such as flush) reduces the runtime by 31 - 56\%.}
    \label{fig:hardware-pipelining}
    \vspace{-0.2cm}
\end{figure}

Fig.~\ref{fig:hardware-pipelining} shows the impact of hardening broadcast signals (Section~\ref{sec:hardware}). In this experiment, all of the software pipelining techniques are applied to isolate the impact of the hardware techniques. As shown in Fig.~\ref{fig:hardware-pipelining}, the runtime is reduced by 31 - 56\%.

\subsection{Pipelining Techniques for Sparse Applications}

\begin{figure}[]
    \vspace{-0.2cm}
    \centering
    \includegraphics[width=0.48\textwidth]{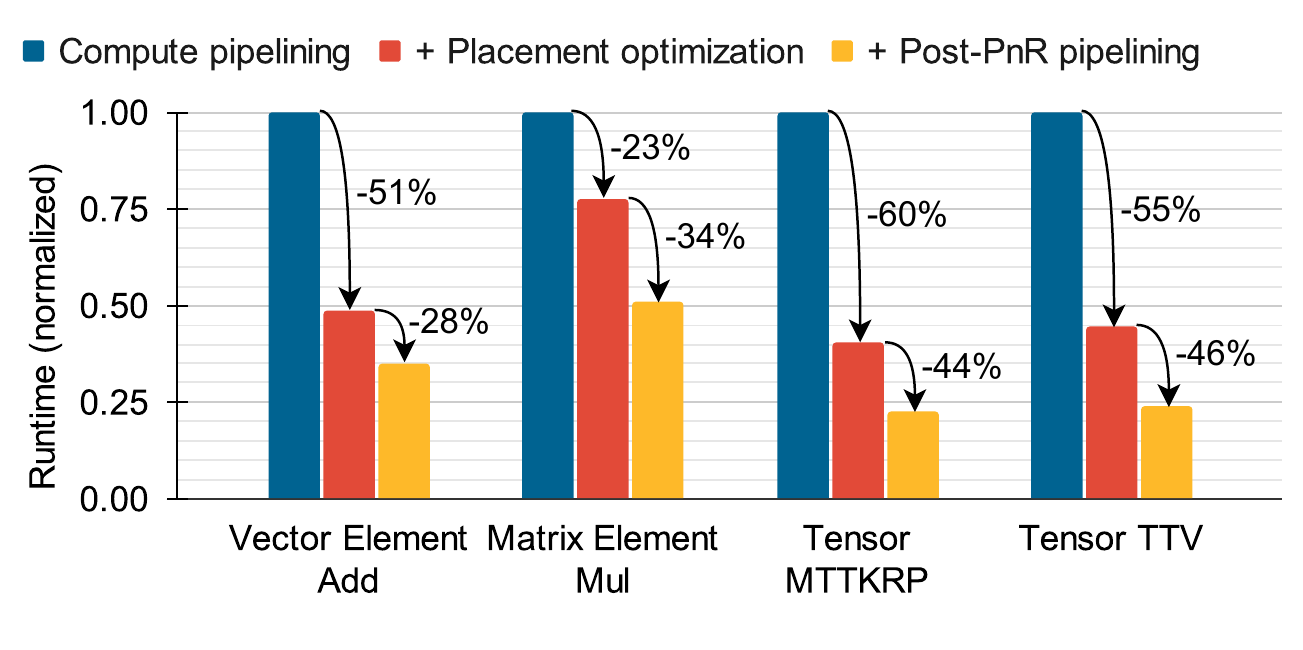} 
    \vspace{-0.4cm}
    \caption{Incremental application of our pipelining techniques to sparse applications. The runtime decreases significantly when placement optimization is applied.}
    \label{fig:sparse-pipelining}
    \vspace{-0.4cm}
\end{figure}

We evaluate the effect of our pipelining techniques from Section~\ref{sec:sparse} on four sparse workloads from \cite{taco}. Note that the sparse applications use FIFOs at the input of every compute unit, so compute pipelining is applied by default by the compiler and cannot be turned off. Additionally, broadcast pipelining and low unrolling duplication had no effect on the frequency so only placement optimization and post-PnR pipelining are evaluated. Fig.~\ref{fig:sparse-pipelining} shows the effect of incrementally applying each technique based on the STA model. Table~\ref{tab:sparse-pipelining} shows the final numbers for maximum frequency, runtime and power comparison for sparse applications, which are verified with SDF-annotated gate-level simulation. As shown in Table~\ref{tab:sparse-pipelining}, the runtime of sparse applications decreases by 29 - 65\% compared to versions with only compute pipelining, and the energy-delay product reduces by 35 - 76\% with our techniques.

\begin{table}\footnotesize
    \caption{Frequency, runtime, and power comparison between compute pipelined and fully pipelined versions of four sparse applications.}\label{tab:sparse-pipelining}
    \vspace{-0.3cm}
    \begin{center}
        \renewcommand{\arraystretch}{1.2}
        \begin{tabular}{llccc}
        \hline\hline
        \multicolumn{2}{c}{Sparse Application}            &   \makecell{Frequency \\ (MHz)} & \makecell{Runtime \\ (µs)} & \makecell{Power \\ (mW)} \\
        \hline\hline
        \multirow{4}{*}{\makecell{Compute \\ Pipelining}}&   Vector Elementwise Add      &   305     &   0.83    &  187 \\
                                                         &   Matrix Elementwise Mul      &   435     &   1.38    &  246 \\
                                                         &   Tensor MTTKRP               &   300     &   33.9    &  194 \\
                                                         &   Tensor TTV                  &   260     &   10.0    &  170 \\
        \hline
        \multirow{4}{*}{\makecell{All \\ Software \\ Pipelining}}  &   Vector Elementwise Add      &    599   & 0.42   &   303   \\
                                                       &   Matrix Elementwise Mul      &    599   & 0.99   &   316   \\
                                                       &   Tensor MTTKRP               &    617   & 14.2   &   320   \\
                                                       &   Tensor TTV                  &    617   & 3.52   &   325   \\
        \hline\hline
        \end{tabular}
    \end{center}
\end{table}

\begin{figure}
    \centering
    \vspace{-0.4cm}
    \hspace*{-0.5cm}%
    \includegraphics[width=0.48\textwidth]{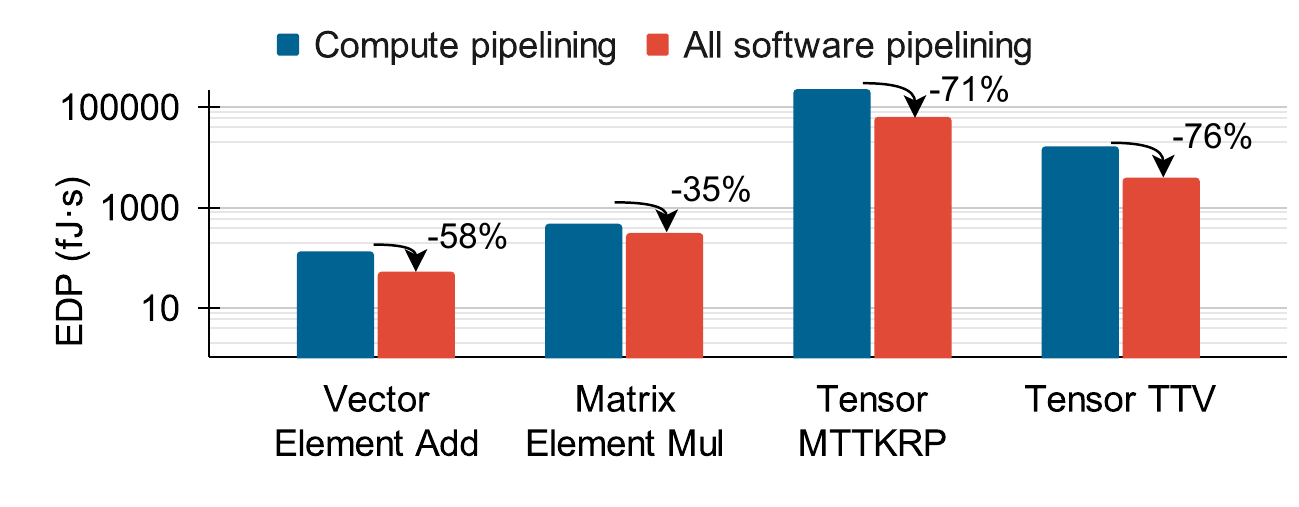} 
    \vspace{-0.4cm}
    \caption{EDP comparison between compute pipelining only and all pipelining applied to sparse applications.}
    \label{fig:EDP-sparse}
    \vspace{-0.4cm}
\end{figure}


\section{Conclusion}
    \vspace{-0.1cm}
Cascade is a CGRA application pipelining toolkit that enables the use of single-cycle multi-hop interconnects while maintaining high performance and energy efficiency. While pipelining tools and techniques have been widely studied for FPGAs and ASICs, existing CGRA compilers either assume expensive exhaustive pipelining or generate applications with poor performance. Cascade enables 7 - 34$\times$ lower critical path delays and 7 - 190$\times$ lower EDP across a variety of dense image processing and machine learning workloads, and 2 - 4.4$\times$ lower critical path delays and 1.5 - 4.2$\times$ lower EDP on sparse workloads, compared to a compiler without pipelining. 
    \vspace{-0.1cm}

\bibliographystyle{IEEETran}
\bibliography{refs}

\end{document}